\documentclass[preprint,12pt,authoryear]{elsarticle}

\usepackage[utf8]{inputenc}
\usepackage[english]{babel}
\usepackage{upgreek}
\usepackage{natbib}
\usepackage{subcaption}
\usepackage{caption}
\usepackage[colorlinks,bookmarksopen,bookmarksnumbered,citecolor=red,urlcolor=red]{hyperref}
\usepackage[hang,flushmargin]{footmisc}
\usepackage{multirow}
\graphicspath{{Figures/}} %
\DeclareGraphicsExtensions{.pdf,.jpeg,.png}
\usepackage{booktabs}
\usepackage{tabularx}
\usepackage[hypcap]{caption}
\usepackage{float}
\usepackage{moreverb}
\usepackage{url}
\usepackage{color}
\newcommand{\fref}[1]{Figure~\ref{#1}}
\newcommand{\tref}[1]{Table~\ref{#1}}

\usepackage{amssymb}
\usepackage{amsmath}
\usepackage{wasysym}

\makeatletter
\newcommand{\mypm}{\mathbin{\mathpalette\@mypm\relax}}
\newcommand{\@mypm}[2]{\ooalign{%
  \raisebox{.5\height}{$#1+$}\cr
  \smash{\raisebox{-.3\height}{$#1-$}}\cr}}
\makeatother
\newcommand{\bpm}{\sbox0{$1$}\sbox2{$\scriptstyle\mypm$}
  \raise\dimexpr(\ht0-\ht2)/2\relax\box2 }
\makeatletter
\newcommand*{\compress}{\@minipagetrue}
\makeatother
\usepackage{array}

\journal{Computer Speech and Language}

\begin{document}

\begin{frontmatter}

\title{Prediction of speech intelligibility with DNN-based performance measures}

\author[aff1]{Angel Mario Castro Martinez}

\affiliation[aff1]{organization={Medical Physics and Cluster of Excellence Hearing4all, Carl von Ossietzky Universit\"at Oldenburg},
            addressline={Ammerl\"ander Heerstrasse 114-118}, 
            city={Oldenburg},
            postcode={26129}, 
            state={Lower Saxony},
            country={Germany}}

\affiliation[aff2]{organization={Communication Acoustics and Cluster of Excellence Hearing4all, Carl von Ossietzky Universit\"at Oldenburg},
            addressline={Ammerl\"ander Heerstrasse 114-118}, 
            city={Oldenburg},
            postcode={26129}, 
            state={Lower Saxony},
            country={Germany}}

\author[aff1]{Constantin Spille}
\author[aff2]{Jana Ro{\ss}bach}
\author[aff1]{Birger Kollmeier}
\author[aff2]{Bernd T. Meyer}

\begin{abstract}
This paper presents a speech intelligibility model based on automatic speech recognition (ASR), combining phoneme probabilities from deep neural networks (DNN) and a performance measure that estimates the word error rate from these probabilities. This model does not require the clean speech reference nor the word labels during testing as the ASR decoding step --- which finds the most likely sequence of words given phoneme posterior probabilities --- is omitted. The model is evaluated via the root-mean-squared error between the predicted and observed speech reception thresholds from eight normal-hearing listeners. The recognition task consists of identifying noisy words from a German matrix sentence test. The speech material was mixed with eight noise maskers covering different modulation types, from speech-shaped stationary noise to a single-talker masker. The prediction performance is compared to five established models and an ASR-model using word labels. Two combinations of features and networks were tested. Both include temporal information either at the feature level (amplitude modulation filterbanks and a feed-forward network) or captured by the architecture (mel-spectrograms and a time-delay deep neural network, TDNN). The TDNN model is on par with the DNN while reducing the number of parameters by a factor of $37$; this optimization allows parallel streams on dedicated hearing aid hardware as a forward-pass can be computed within the 10~ms of each frame. The proposed model performs almost as well as the label-based model and produces more accurate predictions than the baseline models.
\end{abstract}

\begin{graphicalabstract}
\includegraphics[width=\textwidth]{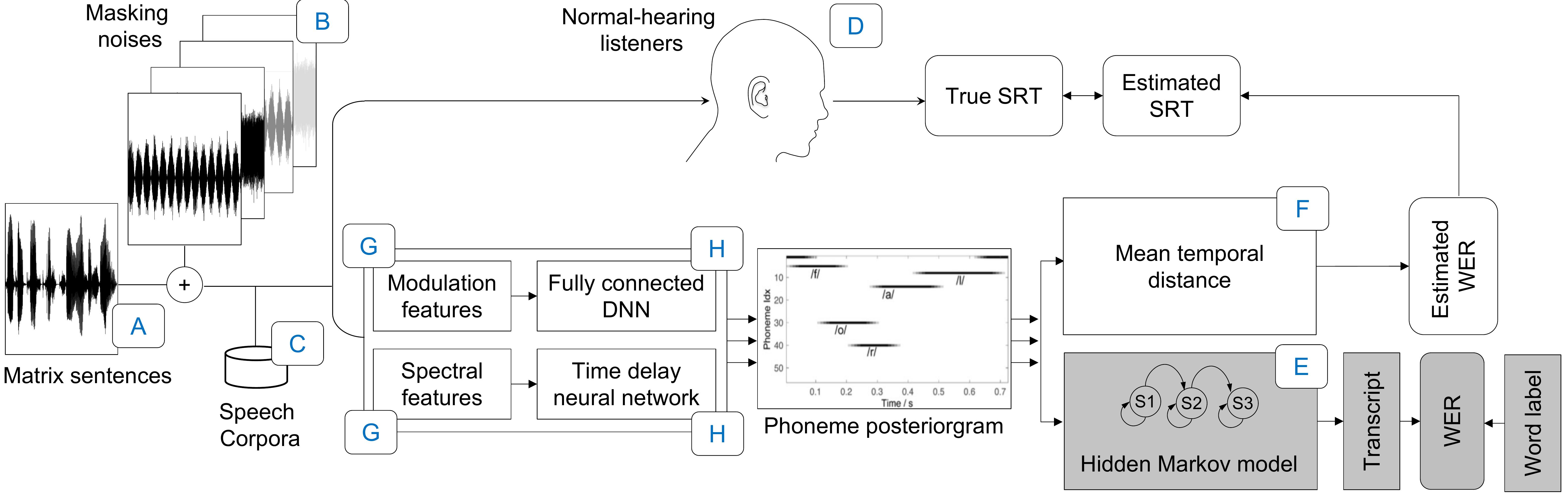}
\end{graphicalabstract}

\begin{highlights}
\item A new speech intelligibility model based on automatic speech recognition (ASR), combining phoneme probabilities from deep neural networks (DNN) and a performance measure that estimates the word error rate from these probabilities.
\item The model does not require the clean speech reference nor the word labels during testing as the ASR decoding step
\item The model allows parallel streams on dedicated hearing aid hardware as a forward-pass can be computed within the 10~ms of each frame and produces more accurate predictions than the baseline models.

\end{highlights}


\end{frontmatter}

\section{Introduction}
\label{sec:intro}
The intelligibility of speech is crucial for our social interaction, and it is an essential measure for a diagnosis of hearing deficits through speech audiometry and the optimization of speech enhancement algorithms in hearing aids or cochlear implants. 
Accurate models that predict speech intelligibility (SI) in the presence of different masking noises are desirable since they can quantify the outcome of such optimization and, therefore, reduce the requirement of SI measurements that are usually time-consuming and costly. 

Several models for SI prediction have been proposed to include the signal-processing strategies of the auditory system. A classic model is the speech-intelligibility index (SII), which is based on the weighted signal-to-noise ratio in separate frequency bands that are combined to estimate SI for longer speech segments (e.g., an utterance) \citep{ANSI1997}.

The extended SII (ESII) \citep{Rhebergen2005} uses the same frequency weighting and covers a short-time analysis of SNRs in each channel, and can take into account the effect of temporal modulations to enable listening-in-the-dips. The short-time objective intelligibility (STOI, \citealp{Taal2011}) quantifies the degradation of speech signals by comparing short-time segments of the original clean speech signal with its degraded counterpart, calculating the correlation of these segments in frequency bands. \citet{Ewert2000a} proposed to include explicit temporal modulation filtering for separate frequency channels, an approach later refined by \citet{Jorgensen2013} by using temporal segments with a variable length that depend on the modulation filter frequency, which resulted in the multi-resolution speech envelope power spectrum model (mr-sEPSM).

\citet{Schubotz2016} compared these models in a study to determine how well they can predict the speech reception threshold (SRT), which is the signal-to-noise ratio (SNR) at which $50$\% of words presented in an acoustically-controlled environment are correctly recognized. In this comparison, mr-sEPSM produced the best average predictions across all maskers that varied from stationary speech-shaped noise to a single background talker in the time-frequency domain. Additionally, this benchmark provides evidence about the influence of amplitude modulation, short-term energy, and informational masking in speech intelligibility.  

Another model that uses auditory periphery modeling to predict the SI is the hearing aid speech perception index (HASPI, \citealp{Kates2014}). It calculates the envelope and the temporal fine structure of the noisy speech signal and the clean speech reference. The coherence between the test and reference signals and the correlation of their envelopes results in the HASPI. This index can predict the SI of normal-hearing listeners and hearing-impaired listeners by including hearing loss into the auditory model; however, the clean speech material is required for the prediction.

An alternative modeling approach combines signal extraction based on auditory principles with pattern matching algorithms borrowed from automatic speech recognition (ASR). \citet{Barker2006} introduced a glimpsing model in which the above-threshold time-frequency patches (glimpses) were used as features for a backend that combines a Gaussian mixture model (GMM) with a hidden Markov model (HMM) to produce a transcript from the input glimpses, which was compared to listener responses.  Recent work by \citet{Tang2016} introduced the binaural distortion-weighted glimpse proportion metric, building upon the previous glimpsing models of monaural signals (\emph{better-ear glimpsing}) and incorporated binaural masking level differences.

\citet{Jurgens2009} combined the output of a model of the internal representation of auditory perception (PEMO, \citealp{Dau1997b}) with the dynamic time warping algorithm, which produced accurate predictions of phoneme recognition in normal-hearing listeners. 

A GMM-HMM approach dubbed Framework for Acoustic Discrimination Experiments (FADE) was proposed by \citet{Schadler2015}. This model produces accurate SRT estimates by retraining a GMM-HMM system at different SNRs and by selecting the model that produces the lowest SRT when using the same training and test sentences.

All previously mentioned models either require separate clean and degraded speech (as in the case of STOI and HASPI) or separate speech and noise signals, either for calculating the frequency-dependent SNR as for the SII, by identifying above-noise speech glimpses as in the model by \citet{Barker2006}, or by using identical underlying speech utterances for training and testing (FADE, \citealp{Schadler2015}). 

The main focus of ASR is to find the most probable uttered word sequence given the acoustic input. This problem can be expressed probabilistically as finding the candidate word sequence $W$ with the highest probability given a set of features $X$ from the observed audio signal.

\begin{equation}\label{eqn:ASR}
    \hat{W} = \underset{w}{\arg\!\max} \; P(W|X)
\end{equation}

Conventionally, the posterior probability is factorized as a product of an acoustic likelihood $P(X|W)$ and a prior model of word sequences $P(W)$ (\emph{language model}). The acoustic likelihood can be obtained by multiplying the conditional distributions of acoustic features (\emph{acoustic model}) with those of a phonetic sequence given a word sequence (\emph{lexicon or pronunciation model}). In simpler terms, the acoustic model predicts the probability of the speech sound given a word sequence, and the language model computes the likelihood of that word sequence.

Motivated by the success of deep learning in ASR \citep{Hinton2012}, \citet{Spille2018} proposed an ASR model that combines a deep neural network (DNN) trained to estimate isolated phoneme probabilities given the acoustic observation with an HMM to consider transition probabilities between phonemes.
 
The predictive power of this model exceeded the four baseline models mentioned above on the dataset collected by \citet{Schubotz2016}. The root-mean-square error (RMSE) between measurement and prediction was 1.8\,dB on average when using multi-condition training and modulation features, which was considerably better than the baseline models RMSEs between 5.6 and 12.5\,dB.

The training and test sets applied to the model are speaker-independent, marking a step towards reference-free SI models, which could serve in assisted hearing. A use case for such a model is the constant monitoring of speech intelligibility in the current acoustic scene and identifying the optimal speech enhancement algorithm for that scene.

However, the previous model uses identical noise signals for training and testing (the effect of disjunct training/testing noises in this model has not yet been explored), and it requires the correct labels of the words in the utterance used as model input. These labels are compared to the transcript produced by the ASR system from which the recognition accuracy is calculated. For online applications of speech intelligibility models, e.g., for real-time comparison and selection of hearing aid and other signal processing algorithms, both the identical noises and the label requirement are essential limitations. Further, it is unclear if the computational demands of the model are compatible with mobile listening devices. 

This paper introduces an SI prediction model that requires neither the speech reference nor the actual labels of the tested utterances. The model follows the DNN-based approach introduced by \citet{Spille2018}. The resulting model is not blind concerning the noise signals, for it follows the training/testing procedure suggested by \citet{Spille2018}, using identical noise signals. The primary enhancement is that we provide a method for \emph{estimating} the word error rate (WER) directly from the phoneme posterior probabilities emitted by the DNN instead of computing the WER between the reference transcript and the one decoded by the HMM.

The WER estimation algorithm is referred to as mean temporal distance (MTD) and was first proposed for estimating error rates of automatic recognizers by analyzing the mean distance of phoneme probability vectors obtained from a neural network over a time interval \citep{Hermansky2013}. Intuitively, the higher the MTD value, the more representative the difference between phonemes contained in the measured time window. In preliminary work, we studied whether MTD was applicable for estimating the WER in SI models, focusing on only four types of masker and one network architecture with comparably high computational demands \citep{Castro2020}. 

When considering applications for models-in-the-loop in hearing aids, the computational complexity becomes quite relevant. In a previous study, we were able to show that fully connected feed-forward DNNs can be used in combination with a hearing aid co-processor \citep{Castro2019}. However, the approach is currently limited to run single models in constrained mobile systems due to the computational complexity of these network architectures. This restriction prevents a simultaneous evaluation of two or more processing strategies (e.g., signal enhancement algorithms in the hearing aid) to select the currently best enhancement.

We also investigate whether the computational complexity can be reduced by replacing the fully-connected network with a time-delay deep neural network (TDNN), recently introduced for efficient ASR \citep{Peddinti2015}. This network architecture models temporal context by introducing input dependencies from previous layers (\emph{delays}); because the input is a concatenated sequence of time frames, these delays can be seen as the input at different points in time.

TDNNs are particularly interesting when modeling speech intelligibility as they cover a broader temporal context than traditional fully connected DNNs. Our previous model included temporal context on feature level using amplitude modulation filters \citep{Moritz2015} and at the input layer by concatenating a given number of frames before and after the current one, thus requiring additional computational resources. Given the number of concatenated features and the input layer size, the overall number of parameters in the network exploded. We explore whether TDNNs can exploit temporal dependencies in the acoustic model more efficiently by taking just some specific context frames instead of all of them. 

Previous research has extensively compared TDNNs with other architectures for acoustic modeling, including DNNs \citep{Peddinti2015}, CNN embeddings \citep{Rownicka2018}, LSTM and similar architectures that implicitly capture temporal dependencies \citep{Huang2019}, and several other architectures using the same framework as our studies \citep{Georgescu2019}. Moreover, feature engineering combined with DNNs has been the subject of our previous studies \citep{Castro2014, Castro2016, Castro2019}. 

\begin{figure*}[htb]
\begin{center}
\includegraphics[width=\textwidth]{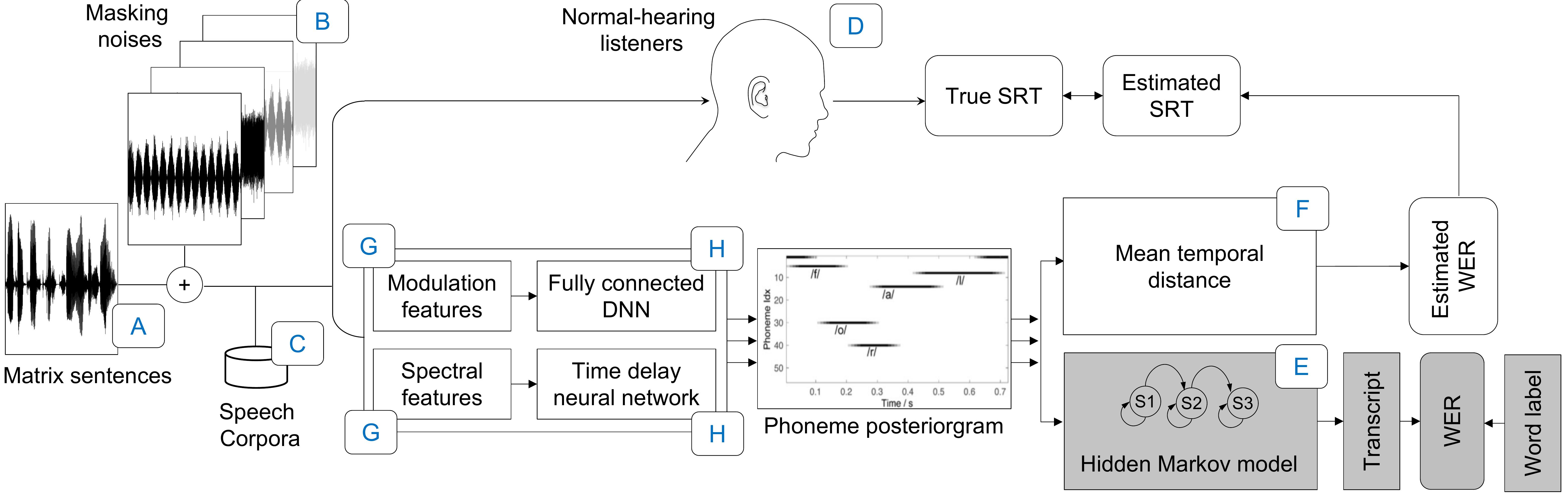}
\caption{Building blocks of the modeling approach: Speech intelligibility in noisy sentences is compared to the estimated SRT. To obtain this estimate, a DNN is trained as part of a standard ASR system and subsequently used to measure the degradation of phoneme representations in noise using a performance measure. From this, the WER of the ASR system is estimated, resulting in the predicted SRT. Blue letters link to the corresponding section of the paper.}
\label{fig:illustration}
\end{center}
\end{figure*}

In summary, the main goals of this paper are to explore ASR-based models of speech intelligibility without the need for transcripts or speech reference in a wide range of maskers and to contrast traditional DNNs with more efficient TDNNs for computational efficiency.

\section{Methods}
\label{sec:methods}
\fref{fig:illustration} serves as a guide through the methods and the proposed model. Starting with the \textit{Oldenburger Satztest} (A) \hyperref[subsec:olsa]{(OLSA)} \citep{Wagener1999}; mixed with the \hyperref[subsec:maskers] {noises maskers}, described in (B). Both building blocks of the \hyperref[subsec:speech]{speech corpora} (C) to conduct the \hyperref[subsec:listening]{listening tests} (D) in the framework for human speech intelligibility developed by \citet{Schubotz2016} and to create the separate datasets for training and testing the proposed model. In previous work, \citet{Spille2018} estimated the SRTs from the target speech transcript, and the WER computed from the output of a regular \hyperref[subsec:asr-si]{hybrid ASR system}. The system components are summarized below (E), together with the modification of using the \hyperref[subsec:m-measure]{M-measure} to drop the requirement of word labels and estimate SRTs directly from the output of the acoustic model (F). The input \hyperref[subsec:features]{features} and the \hyperref[subsec:net]{acoustic models} are described in (G) and (H).

\subsection{Matrix Test (A)}
\label{subsec:olsa}
The \textit{Oldenburger Satztest} (OLSA) \citep{Wagener1999} is a matrix sentence test. It consists of $120$ utterances produced by one speaker. Target sentences derived from a vocabulary of $50$ words equally divided into five categories. For a review of matrix tests in several languages, please refer to \citep{Kollmeier2015}.

Each sentence is comprised of five words following the fixed structure: \textless name\textgreater\ \textless verb\textgreater\ \textless number\textgreater\ \textless adjective\textgreater\ \textless object\textgreater\, e.g. \textit{"Peter verkauft sechs nasse Tassen"} ("Peter sells six wet cups"). Despite being grammatically correct, these sentences have no semantic context. Moreover, all combinations of words from each of the five categories can occur; therefore, predicting a sentence from previous ones is highly unlikely.

\subsection{Noise Maskers (B)}
\label{subsec:maskers}
In the study carried out by \citet{Schubotz2016}, a set of eight background maskers was created to evaluate the effect of energetic, amplitude modulation and informational masking on speech intelligibility.
We took this benchmark to evaluate our speech intelligibility prediction model. 

The maskers have different spectro-temporal characteristics split into two groups according to the type of noise derived from:
First, a stationary speech-shaped noise (SSN) with the same long-term spectrum as the International Speech Test Signal \citep[ISTS;][]{Holube2010} was used. Second, a sinusoidally amplitude-modulated SSN (SAM-SSN) was produced by adding an $8$\,Hz temporal modulation. The third masker was generated by multiplying the Hilbert envelope of a broadband speech signal with the SSN (BB-SSN). For the fourth, named across-frequency shifted SNN (AFS-SSN), the SSN was filtered in $32$ frequency channels; subsequently, every four adjacent channels were multiplied with a different random time section of the Hilbert envelope used for BB-SSN\footnote{resulting in eight different adjacent modulation bands}.

Four additional speech-like maskers come from two unmodified speech signals for the second group: the ISTS and a single talker (ST) and the noise-vocoded versions (NV-ISTS and NV-ST, respectively), removing the speech fine-structure. The speaker in ST is not the same as the one uttering the target signal, nor is any of the $20$ used for ASR training material described below.

To test the influence of same- or different-gender maskers, all maskers have a male and female version to match the long-term spectrum of the respective gender. Since the original ISTS contains female voices only, \citet{Schubotz2016} produced a male version of ISTS via the STRAIGHT algorithm \citep{Kawahara2008} to match its long-term spectrum. The female version of ISTS served as the basis for six out of eight of the maskers. 

\tref{tab:maskers}  contains a summary of the resulting eight maskers for each gender.

 \begin{table}
 \footnotesize
 \begin{tabularx}{\textwidth}{p{3.0cm}p{2.7cm}p{3.2cm}p{3.2cm}}
 \multicolumn{4}{l}{Speech-shaped noise based maskers} \\
 \vspace{0.1cm} \\
 SSN & SAM-SSN & BB-SSN & AFS-SSN \\
 \toprule
 stationary\,speech-shaped noise & sinusoidally amplitude-modulated SSN & SSN modulated with Hilbert envelope of a broad-band speech signal & across-frequency shifted\,modulated SSN \\
  \vspace{0.2cm} \\
 \multicolumn{4}{l}{Speech-like maskers} \\
 \vspace{0.1cm} \\
 ISTS & NV-ISTS & ST & NV-ST \\ 
 \toprule
 International\,speech test signal & noise-vocoded ISTS & single talker & noise-vocoded single talker \\ 
 \end{tabularx}
 \caption{\label{tab:maskers} All maskers based on speech-shaped noise and all speech-like maskers. For each masker, a male and female version was created.}
\end{table}

\subsection{Speech Corpora (C)}
\label{subsec:speech}

An in-house corpus of $10$ hours of speech from $20$  speakers ($10$ male, $10$ female) with the syntactical structure of OLSA \citep{Meyer2015}, which was selected as a starting point to train both ASR systems; sentences from the original OLSA speaker were not contained in the training set.

Two training sets were created, using statistics estimated from male and female voices to generate noise signals referred to as male and female noise, as described below. Both sets comprise the clean data mixed with random parts of each of the eight different maskers at random uniformly distributed SNRs ranging from $-10\,dB$ to $20$\,dB, resulting in $80$\,h of speech material per set. Therefore, training and test sets contain information about the same noise signals, i.e., the model is not blind to the noise, which hinders its use as a model-in-the-loop (as also pointed out in the discussion). 

The ASR training was based on $10$ hours of German matrix sentences collected from $20$ speakers. Additive noise was used at an SNR range from $-10$ to $20$\,dB, resulting in an extended training set of $80$ hours \footnote{The noises are 60 seconds long, and random segments of the same signals with different random seeds were used for training and testing}. ASR testing was carried out with only $120$ matrix sentences from one speaker (clean data). 

Finally, to sample the entire psychometric function of the ASR system, a test set was created by mixing random parts of the respective masker with eight different random sentences from the original OLSA dataset for each of the $400$ SNR values uniformly distributed between $-30$\,dB and $20$\,dB.

 \begin{table*}
 \footnotesize
 \begin{tabularx}{\textwidth}{p{1.0cm}p{3.5cm}p{3.0cm}p{4.5cm}}
  & ASR training set & ASR test set & listening experiments \\
 \toprule
 \multirow{2}{*}{Clean} & 20 speakers & 1 speaker & 1 speaker \\
            & 10\,h & 120 sentences & 120 sentences \\
 \midrule
 \multirow{2}{*}{Noisy} & 80\,h multicondition set & 8 different random sentences per SNR &  adaptive SRT measurement \\
            & SNRs [$-10$\,dB - $20$\,dB] & 400 random SNRs [$-30$\,dB - $20$\,dB] & 20 sentences per masker and listener for SRT$_{50}$ \& SRT$_{80}$  \\
 \end{tabularx}
 \caption{\label{tab:speech} Overview of the different speech datasets for training, testing of the ASR systems and the listening experiments}
\end{table*}

\subsection{Listening Tests (D)}
\label{subsec:listening}
To test the performance of the proposed speech intelligibility predictor, we compare the results from the listening tests performed in \citep{Schubotz2016}. These experiments consisted of characterizing speech intelligibility as a function of the speech reception threshold extracted from the adaptive procedure proposed by \citet{Brand2002}. 

Eight normal-hearing participants (ages between 23-34) participated who were not previously exposed to the speech task; their hearing thresholds for pure tones did not exceed $20$\,dB at frequencies between $125$\,Hz and $8$\,kHz. During testing, the participants attended  $20$ OLSA sentences (as described in the \hyperlink{subsec:olsa}{Matrix Test} section below) with an initial SNR of $0$\,dB; then, the SNR varied depending on the intelligibility measurement of the previous sentence. 
The procedure is set to determine the SNRs at which listeners correctly understand $50$\% (SRT) and $80$\% (SRT$_{80}$) of presented words. Each SRT resulted from a different set of sentences; in other words, each participant listened to $40$ sentences per noise condition.
Finally, the SRTs were averaged across the listeners to obtain the final SRT and SRT$_{80}$ values, which are used to trace the psychometric function of the listeners (described entirely by the SRT and its slope). The psychometric function slope was estimated via a maximum-likelihood estimator \citep{Brand2002} with the $40$ responses for each listener and masker.

\section{Speech Intelligibility Model}
\label{sec:si-model}
\subsection{ASR-based Model of Speech Intelligibility (E)}
\label{subsec:asr-si}
Speech intelligibility models based on ASR work on the premise of comparing human and automatic speech recognition capabilities. Traditionally, the two components of a speech recognizer, namely the acoustic and language models, are trained separately. In this work, we require both for the training phase; when inferring, we decouple the latter because the output of the acoustic model is directly used instead of the decoded utterance.

In the label-based ASR approach, \citet{Spille2018} trained the acoustic model on speech files mixed with different maskers at various SNRs. A fully connected feed-forward DNN was used to map the acoustic features to posterior probabilities of context-dependent triphones. The time sequence of these probabilities was decoded using an HMM (three states for modeling phonemes and five for silence) to obtain a transcript of the utterance. 

This transcript was compared to the ground truth labels to obtain the word error rate (WER) from this sentence. By using utterances at various SNRs, a broad range of the corresponding WER estimations was obtained. 

These pairs of points were fitted to a psychometric function, as described by \citet{Wagener1999}. 
Word recognition accuracies were taken instead of the WER to be consistent with the form of the psychometric function $f(x)$, which traditionally measures human speech intelligibility as a function of the signal-to-noise ratio $x$ in $dB$.

 \begin{equation}\label{eq:psychometric}
     f(x) = \frac{1}{1 + e^{4s(L_{50} - x)}}\ ~
 \end{equation}

$L_{50}$ is the SRT and $s$, the fitted slope parameter. SRT values served as a mean intelligibility predictor and were compared to SRTs obtained in listening experiments. 

The proposed model differs from this approach by using the HMM during the training procedure only and omitting the HMM (or any other language model) when predictions were obtained. Instead, as described in the next section, a performance measure quantifies the degradation of the phoneme posteriorgrams. This measure could be informative about the WER and hence, about the SRT too.

We hypothesize the resulting measure should be sensitive to the SNR and show a similar sensitivity to masking noises similar to human listeners as long as the training data is sufficient.

\subsection{Estimating the Word Error Rate (F)}
\label{subsec:m-measure}
The WER estimation is based on quantifying the degradation of phoneme probabilities obtained from a DNN. We chose the mean temporal distance (MTD, also referred to as M-measure) \citep{Hermansky2013}, which considers the distance of phoneme vectors and averages this distance. 

\begin{figure*}[htb]
\begin{center}
\includegraphics[width=\textwidth]{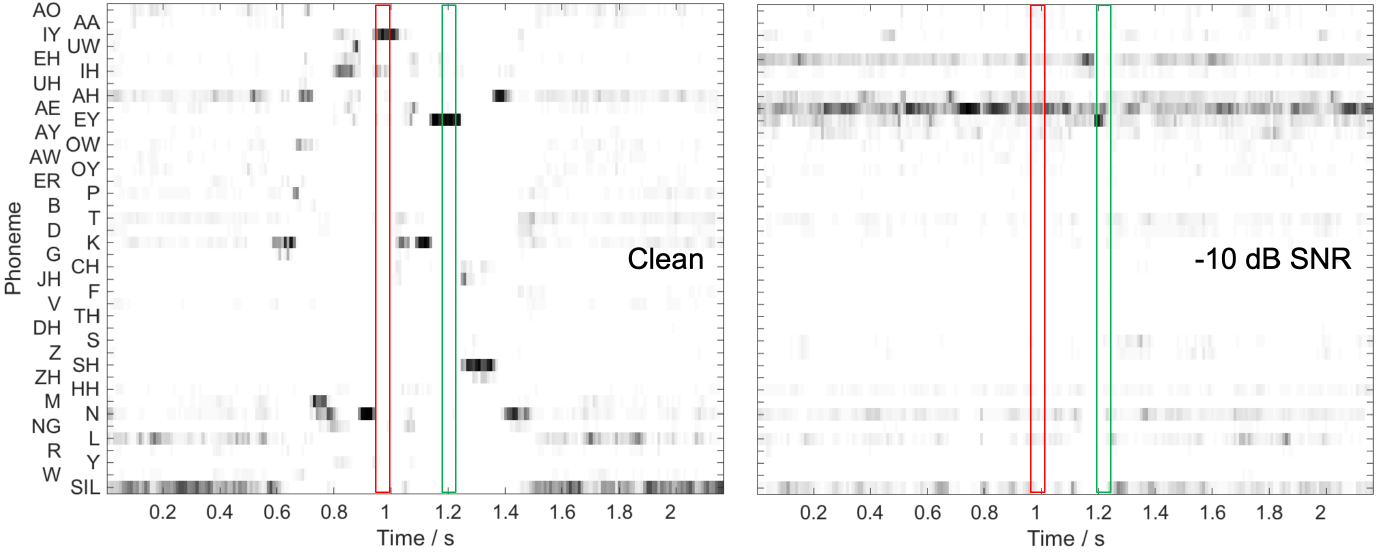}
\caption{Two posteriorgrams at different SNRs showing the temporal smearing effect captured by the mean temporal distance. Two frames are selected across $200$\,ms in both conditions, representing how distant frames become more similar in acoustically challenging scenes.}
\label{fig:mtd-smearing}
\end{center}
\end{figure*}

The underlying idea is that acoustically challenging conditions can have a temporal smearing effect on phoneme representations, i.e., the phoneme vectors become more similar. This effect is shown in \fref{fig:mtd-smearing}, where two different frames are selected across a $200$\,ms window in a clean posteriorgram and one at $-10$\,dB SNR. We can observe how the noisier the setup, the more similar distant phonemes become. Acoustically optimal conditions produce very distinct phoneme activations, which become more distant in vector space. This distance is captured by the entropy-based divergence that averaged over phoneme vectors in the interval from $50$ to $800$\,ms. 

We propose overcoming the need for the actual transcript of the reference utterance via performance measures, particularly the M-measure. This measure introduced by \citet{Hermansky2013} indicates the mean distance between probability estimations (in this case, the phoneme posteriorgram output by the acoustic model) over a specific time interval. The central intuition for using posteriorgrams to reflect speech quality is that their distribution relates to the type of noise as well as to the SNR; in clean conditions, the peaks are sparse and isolated, clearly favoring the predicted phoneme, whereas those peaks flatten when noise is added to the system. This effect is critical for the M-measure since similar phoneme vectors, such as coarticulation, result in smaller distances, whereas distinct vectors result in greater distances.

The M-measure accumulates the average divergences of two phoneme posterior vectors $\mathbf{p}_{t-\Delta t}$ and $\mathbf{p}_t$ separated by a time interval of $\Delta t$ and is defined as
\begin{equation}\label{eqn:MTD}
    \mathcal{M}(\Delta t) = \frac{1}{T-\Delta t} \sum^T_{t=\Delta t} \mathcal{D}(\mathbf{p}_{t-\Delta  t},\mathbf{p}_t)\ ~
\end{equation}

Where $T$ is the duration of the analyzed representation, in this case, a portion of the posteriorgram. The symmetric Kullback-Leibler divergence \citep{Kullback1951} is used as distance measure $\mathcal{D}$ between phoneme posterior vectors $\mathbf{p}_{t-\Delta t}$ and $\mathbf{p}_t$.
 
\begin{equation}\label{eqn:kldiv}
    \mathcal{D}(\mathbf{p,q}) = \sum_{k=0}^Kp^{(k)}\log{\frac{p^{(k)}}{q^{(k)}}} + \sum_{k=0}^Kq^{(k)}\log{\frac{q^{(k)}}{p^{(k)}}}
\end{equation}

As defined above, $p^{(k)}$ is the $k$-th element of the posterior vector $\mathbf{p}\in\mathbb{R}^k$. 
 
We considered $16$ values of $\Delta t$ per utterance; from $50$ to $800$\,ms in steps of 50\,ms\footnote{The upper limit was considered to capture coarticulation effects as suggested in \citep{Hermansky2013}}. For short $\Delta t$ time spans, divergences are small, indicating neighboring frames often correspond to the same phoneme. Over time, the value increases to a point at which both vectors $\mathbf{p}$, and $\mathbf{q}$ come from different coarticulation patterns, and the curve saturates. 
 
As the acoustic model is trained to produce triphone posteriorgrams, to be decoded by the language model when performing automatic speech recognition, we \textit{grouped} those activations to obtain monophone posteriorgrams. 

It is possible to cluster triphones by mapping each transition as a branch of a decision tree, which roots correspond to the central phoneme of the triphone. Monophone posteriorgrams of $42$ dimensions were obtained by adding the corresponding activations, thus maintaining the distribution\footnote{The list of phonemes are compliant with the ARPABET phonetic transcription code and include the /SIL/ and /NSN/ classes for silence and noise respectively.}.

Monophone posteriorgrams yield M-measure values comparable to those obtained with the triphone equivalents at a lower computational cost without constraining the acoustic model of the temporal context if trained to produce monophones directly.

In our previous study, \citep{Castro2019}, we established the correlation between WER and the M-measure. In this work, we leverage this property, but the estimator ought to be in the same domain as the word recognition accuracy to estimate the SRTs and produce similar psychometric curves as human listeners. Given the non-linearity introduced by the M-measure, a mapping function is required to estimate WERs, which was fitted from the observed relationship between the WER estimations and the corresponding M-Measure values. 

The function used to map the M-measure to WER depends on the acoustic model and decays exponentially according to the following equation:

\begin{equation}\label{eqn:mtd2wer}
    WER(\mathcal{M}) = Ae^{k\mathcal{M}}. 
\end{equation}

The initial value $A$ and the decay rate $k$ were calculated on an independent validation set comprising utterances spoken by a speaker not included in the training set mixed with the same noise maskers (and different random segments) described in the following section\footnote{The values obtained were $A=289.93, k=0.213$.}. Additionally, an upper boundary of $100$ (the highest possible error rate) was imposed by replacing all higher values with $100$ \footnote{The max value was $103.68$}.

\subsection{Features (G)}
\label{subsec:features}
As a starting point, $40$-dimensional log mel frequency spectral coefficients (MFSC)\footnote{In other studies, these coefficients are often referred to as Fbank features. In contrast with standard MFCC features, the last DCT (as inverse Fourier transform) is not applied; hence the domain is the Mel spectrum of the spoken utterance instead of the cepstrum}, calculated from the spoken utterance by transforming the digital signal into a spectrogram via a short-time Fourier transform (STFT) with a shift of 10\,ms and a window length of 25\,ms. The $40$ filters equidistant on a mel scale are applied to each STFT frame, which results in a higher resolution for lower frequencies and a lower resolution for high frequencies. 

The amplitude modulation features were derived from these coefficients by applying an amplitude modulation filter bank (AMFB), designed by \citet{Moritz2015} and refined for optimal automatic speech recognition performance \citep{Moritz2016}. This modulation filterbank leverages physiological studies on the human auditory system and psychoacoustic results \citep{Drullman1994a, Drullman1994b, Elliott2009}, suggesting that most relevant linguistic information is coded in amplitude fluctuations. The principle behind these filters is to mimic the processing strategies of the auditory system by processing the temporal dynamics of speech. 

In \citep{Moritz2015}, their AMFB adopted the time-frequency resolution of amplitude modulation filters from the psychoacoustic model of \citep{Dau1997a}. In contrast, the latest iteration presented in \citep{Moritz2016} optimized these parameters based on multiple speech recognition tasks, such as Aurora-4 \citep{Parihar2004} and the recognition of data from the Librispeech corpus \citep{Panayotov2015}. 

The following functions describe a single amplitude modulation filter $q$, which is a complex exponential function $s$ windowed with zero-phase Hann envelope $h$. 
 \begin{align}
    s(n_c) &= \exp \left(i\omega (n_c - n_0)\right) \\
    q(n_c) &= s(n_c) \cdot h(n_c) \\
    h(n_c) &= \frac{1}{2} + \frac{1}{2}\cdot \cos\left(\frac{2\pi(n_c-n_0)}{W_{n_c} + 1}\right)
 \label{eq:am-filters}
 \end{align}

$n_c$ denotes the frame index, $\omega$ is the radian frequency, and $W_{n_c}$ is the window length of the envelope with the central frame index $n_0$.
The optimized AMFB is integrated by five filters characterized by a cubic relation of their modulation frequency and ($-3$\,dB) bandwidth with center modulation frequencies of $0$, $5.5$, $10.15$, $15.91$, and $27.03$\,Hz.

AMFB features are generated by convolving each frequency channel independently, with each of the five amplitude modulation filters $q$. The resulting real and imaginary parts are concatenated to form the feature vector, except for the DC filter, of which only the real part is used as its imaginary part is zero. Additionally, the mean of the real AMFB parts is subtracted to ensure DC-free filter outputs and a notch at 0 Hz in the frequency response. Finally, AMFB features are mean and variance normalized on an utterance basis. The feature vector at a given frame has $360$ dimensions ($9$ filters convolved with $40$ frequency channels), then spliced $\pm 5$ frames, producing an input layer final size of $3960$.
 
\subsection{Acoustic Modeling Deep Neural Networks (H)}
\label{subsec:net}
Two different deep neural network architectures were trained to map the acoustic observations to phoneme probabilities using the Kaldi toolkit\footnote{The ASR was implemented using the \href{http://kaldi-asr.org}{Kaldi speech recognition toolkit} \citep{Povey2011}. }. These models incorporate temporal context either on feature level (using AMFB features and concatenating contiguous frames) or by the network topology (in case of the time-delay network); the latter is explored since it is computationally more efficient, which could be relevant for future mobile applications of the modeling approach.

Both networks were trained to classify context-dependent triphones; every phone is modeled with three HMM states except for silence, which uses five states. 

The first architecture was a fully connected network (referred to as merely DNN) with six hidden layers and $2048$ (sigmoid) hidden units. We selected the fully connected network from \citep{Spille2018} to appropriately compare this work and the previous SI prediction model. The second consists of a time-delay neural network (TDNN) \citep{Peddinti2015} with four hidden layers and a temporal context of $\pm 2$ frames at the input layer. The TDNN was designed to have a maximum of 1M parameters with upper boundary constraints inspired by our previous work \citep{Castro2019}. This decision would allow the forward pass to be computable within the target processor introduced by \citet{Hartig2014} while covering a similar initial temporal context to the DNN model, smaller shifts from lower hidden layers, and broader shifts in the higher layers.

The DNN was trained with AMFB features processed with modulation filters ranging from $5$ to $20$\,Hz \citep{Moritz2015}. They were chosen since the explicit coding of temporal modulations increased model performance, especially for the AFS-SSN masker previously \citep{Spille2018}. 

The weights were initialized with a stack of Restricted Boltzmann Machines trained unsupervised one layer at the time via contrastive divergence \citep{Hinton2010} and then were fine-tuned to minimize the cross-entropy between output and target vectors. 

The training was performed in up to $20$ epochs (stopping when the relative improvement was lower than $0.001$). The starting learning rate was $0.008$ (halving it every time the relative improvement was lower than $0.01$); these changes were the only modifications to an otherwise vanilla stochastic gradient descent optimizer. A soft-max layer of $1877$ (corresponding to the number of pruned triphones after training) units was attached to the output to produce the most likely posterior probabilities of each context-dependent HMM state. Including the output layer, the total number of weights is $37,127,000$ for this architecture.

The TDNN, in contrast, was trained on $23$-dimensional MFSC\footnote{We found our TDNN does not significantly perform better when adding more mel-frequency spectral bands.}, each frame is taken together with the preceding and following two to provide the temporal context required by the design of this network, resulting in an input layer of $115$ dimensions computed each forward pass. The TDNN contains six composed layers in the form: affine - [batch norm] - ReLU, meaning an affine layer followed by a batch normalization operation \citep{Ioffe2015} with a rectified linear unit activation function \citep{Nair2010} using L2-regularization of $0.05$ on each layer to correct for effects related to the \textit{generalized model averaging} technique described in \citep{Povey2014}. 

The first four layers have an output of $250$ units, and the two remaining have $120$ to adjust for the increased temporal window fed to each of the neurons in the upper layers, enforcing distillation of information and capturing higher-level semantics. Training followed the approach based on natural gradient-stabilized SGD parallelized over 8 GPUs starting with a minibatch of $256$ and halving its size towards the end of the training. The soft-max output layer has $1657$ units giving a total number of weights is $926,187$ for this architecture\footnote{The number of triphones in each architecture is different due to the pruning process of the trained HMM states.}.

Each layer in a TDNN operates at a different temporal context, which increases with every upper layer. The required input frames of each layer parametrize the output activation at every time step. In this case, the layerwise context indicated as the pair of context frames subsampled left and right from the preceding layer is as follows: for the input layer {$-2$, $-1$, $0$, $1$, $2$},  the second to fifth layer {$-1$, $1$}  and the sixth layer {$-2$, $2$}. The total left and right context are from $-8$ until $+8$ frames\footnote{This context effectively means the network can compute up to eight frames in the past and future at any given time.}.
 
The training recipe detailed in \citep{Zhang2014} comprises greedy layerwise supervised training as initialization technique; preconditioned stochastic gradient descent updates, as opposed to fixed learning rate; exponential schedule and \textit{mixing-up}, instead of linear decay.

In principle, the speech recognizer could have been trained on any German corpus containing the same phonemes and vocabulary present in OLSA because the test sentences were produced randomly with a fixed syntax. Given the nature of the task, we created a deterministic language model (including only the words in the matrix sentence test) as we observed no benefit from learning one during training or using a stochastic model.

\section{Results}
\label{sec:results}

\begin{figure}[H]
\begin{subfigure}{\textwidth}
  \centering
  \caption{Female TDNN model for speech-shaped noise maskers}
  \includegraphics{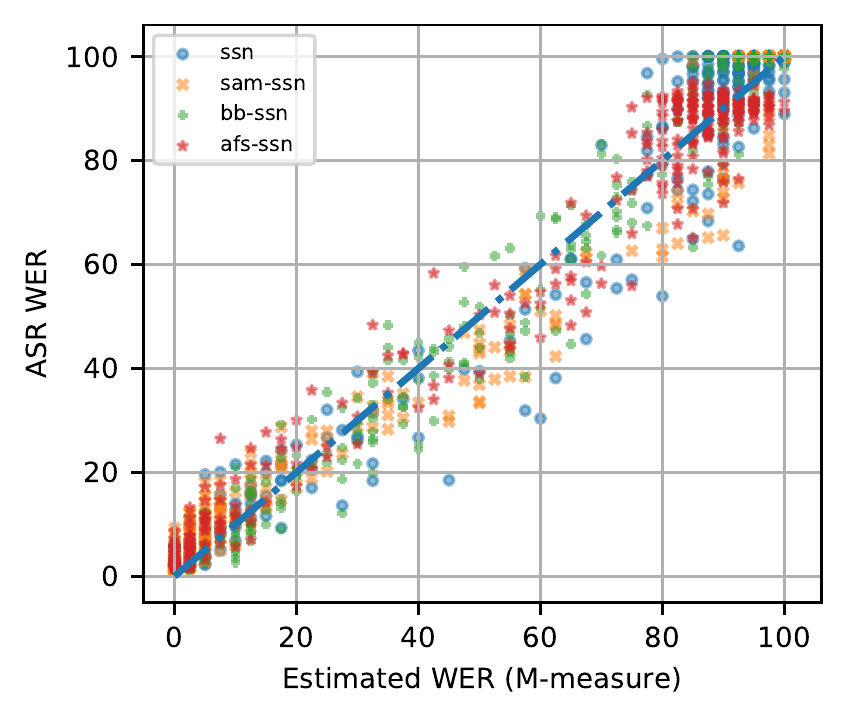}
\end{subfigure}
\begin{subfigure}{\textwidth}
  \centering
  \caption{Female TDNN model for speech-like maskers}
  \includegraphics{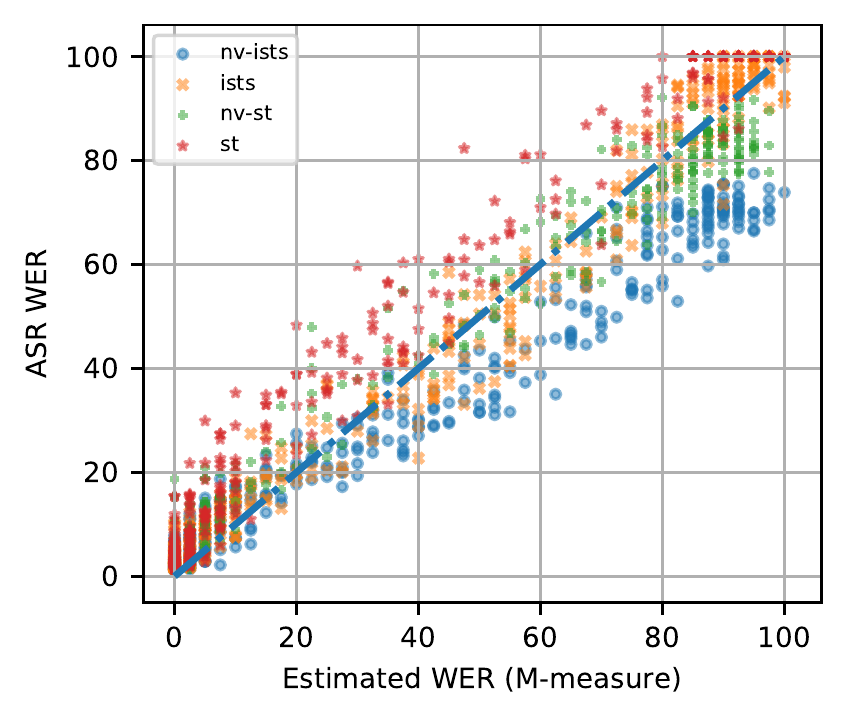}
\end{subfigure}
\caption{Relation of estimated and measured WER for the female set up using the TDNN system trained on MFSC. (a) covers the SSN-based maskers and (b) the four speech-like maskers based on ISTS and ST.}
\label{fig:asr_wer_vs_mtd_estimate}
\end{figure}

Because the modeling approach presented in this paper is based on estimating the WER (cf. \fref{fig:illustration}), we first analyze if the error rate from the ASR is related to the predicted one based on the M-measure (\fref{fig:asr_wer_vs_mtd_estimate}), where each data point corresponds to the error rate for eight matrix sentences, as described in the methods section. 

Panels (a) and (b) show the relation between the measured WER from the ASR and the estimated from phoneme probabilities using Mel-spectrograms and the TDNN architecture. This system performed quite similarly in terms of correlation and root-mean-square error (RMSE) ($r = 0.98$, $RMSE = 7.71$) to the one using of amplitude modulation filterbanks with a fully-connected DNN ($r=0.97$, $RMSE = 9.13$). As the TDNN approach has a lower computational footprint, relevant in the context of mobile applications (cf. \hyperref[subsec:net]{computational complexity}), the following results are based on the TDNN system except for the results summary, which also covers the AMFB-DNN results.

In \fref{fig:asr_wer_vs_mtd_estimate} (a), we observed a mild overestimation of the WER within $40 - 80$\,\%, for the maskers based on SSN, from now on referred to as speech-shaped. In contrast, panel (b) exhibits a compensated under- and overestimation for the NV-ISTS and ST conditions, respectively, allowing the mapping function to generalize to different acoustic conditions. In both panels, we observe a clear relationship between estimated and measured WER for each masker. Additionally, the mapping is most sensitive at lower word error rates as the mapping function is a decaying exponential constrained to an upper boundary of $100$.

When comparing the WER-based and the estimated WER-based SI models, the overestimated WER values result in a lower SRT by the proposed model; the most noticeable cases are SSN ($RMSE = 7.15$) and NV-ISTS ($RMSE = 11.86$) in \fref{fig:asr_wer_vs_mtd_estimate}. Conversely, the shift to a higher is a consequence of underestimating the WER, as observed in the ST ($RMSE = 9.29$) masker. 

\begin{figure}[H]
\begin{center}
\includegraphics[width=\textwidth]{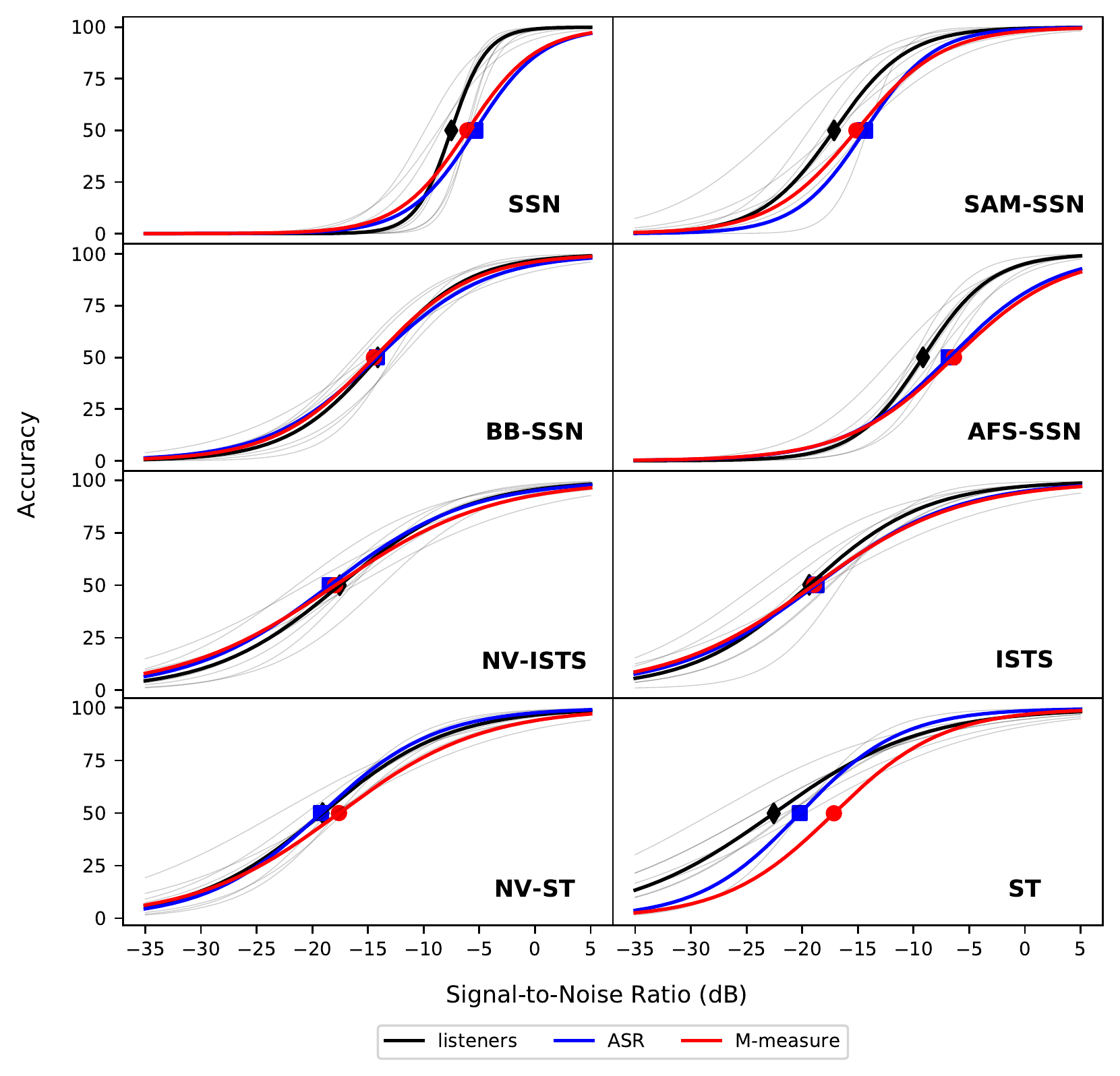}
\caption{Psychometric functions or NH listeners (mean in black, individual curves in gray), the ASR-based model that used \emph{a priori} knowledge in the form of transcripts (blue) and the proposed model that estimates the SRT from phoneme probabilities without using the transcript (red).}
\label{fig:psyfun_dnn}
\end{center}
\end{figure}

To quantify the model performance, we compare the psychometric functions of the listeners to the approach using ASR-generated transcripts and the proposed approach (\fref{fig:psyfun_dnn}). As observed, the proposed model closely matches the human measured psychometric functions (gray lines in the figure).

The differences measured in RMSE between predicted and observed SRTs, calculated for five baseline models, are shown in \tref{tab:RMSE}. In part A, the average SRTs of listeners are compared with those yielded by the proposed model, referred to as AMFB-M-measure. The models trained on the female noise maskers are matched to the corresponding female normal-hearing SRTs; likewise, the male results were compared to a model trained on male noise maskers. 

In both setups, SSN and SAM-SSN, the predicted SNRs were lower than the observed ones, whereas the opposite behavior occurs with the BB-SSN and AFS-SSN noise maskers.
\begin{table*}[!t]
\footnotesize
\begin{tabularx}{\textwidth}{lcccccccc}
\multicolumn{9}{l}{(A) Difference between average SRTs of human listeners and proposed models in dB} \vspace{0.1cm}    \\
                                    &    SSN        &    SAM-SSN    &    BB-SSN    &    AFS-SSN    &    NV-ISTS    &    ISTS    &    NV-ST    &    ST    \\ 
\toprule
\female (LT) &    -7.5       &    -17.1        &    -14.1    &    -9.2    &    -17.5    &    -19.4    &    -19.1    &    -22.6    \\ 
DNN    &    -2.0       &	 -2.9         &    -0.2     &     0.1    &	  -1.1     &    -0.4     &      3.3	   &      3.4  \\
TDNN   &     1.4	    &     2.6	      &    -0.3 	&     2.8    &    -0.4     &	 0.5	 &      1.5    &	  5.6  \\
\midrule
\male (LT)   &    -8.2        &    -17.7    &    -14.9    &    -9.3    &    -18.7    &    -16.3    &    -18.0    &    -15.4    \\
DNN    &    -1.8	     &    -0.8     &	  1.9	 &     2.1    & 	 4.1	&      1.9    &      3.1    &	 -1.7   \\
TDNN   &     0.5        &     2.6     &	 -0.2	 &     4.1	  &      2.2    &  	  -0.3    &	     2.0    &	 -0.8   \\   
\bottomrule
\vspace{0.2cm}
\end{tabularx}
\begin{tabularx}{\textwidth}{lccccc|c|cc}
\multicolumn{9}{l}{(B) Root-mean-squared error in dB of observed and predicted SRTs} \vspace{0.1cm} \\
        &    \multicolumn{5}{c}{baseline models}    &    \multicolumn{1}{c}{Spille et al} & \multicolumn{2}{c}{M-measure}    \\ 
        &    SII     &    ESII    &    STOI    &    mr-sEPSM &   HASPI   &   AMFB &   AMFB &   MFSC \\ 
\toprule
male      &    7.9  &    5.6    &    9.2    &    3.5    &    12.6   &    1.9    &   2.4    &   2.0 \\
female    &    9.4  &    7.6    &    9.7    &    7.8    &    12.4   &    1.7    &   2.1    &   2.5 \\
\midrule
avg.    &    8.6    &    6.6    &    9.5    &    5.6    &   12.5    &    1.8    &   2.2    &   2.3 \\
\bottomrule
\vspace{0.1cm}
\end{tabularx}
\caption{\label{tab:wiebke}(A) SRTs from normal-hearing listening tests (LT) while hearing female and male target speech and the corresponding SRT prediction for the proposed ASR-based models DNN with AMFB features and TDNN with MFSC. The name reflects the architecture and the feature used to train the model (B) SRT  prediction error for baseline models, the label-based previous model and the proposed approach.}
\label{tab:RMSE}
\end{table*}

We compute the RMSE between observed and predicted SRTs to measure the precision of the proposed model in all noise conditions shown in part B of \tref{tab:RMSE}. Among the baseline models, mr-sEPSM yields the lowest error with an average of $5.6$, despite the mr-sEPSM error of the female version was $7.8$, more than twice the male error of $3.5$.

Both ASR-based models were trained with gender-independent speech and maskers; thus, they produce consistent predictions for male and female maskers.

DNN-based models show lower RMSE than the previous models. The model from \citet{Spille2018}, with an average RMSE of $1.8$\,dB, remains the closest to the human observed SRTs For SII, STOI, HASPI, and our proposed model, the error difference between the genders is relatively tiny. The $0.5$ difference between the male and female versions of the TDNN model highlights the most deviant prediction error of $5.6$\,dB for the ST female condition among all proposed model noises.

Finally, a preliminary analysis of the computational complexity of both proposed setups, i.e., the DNN using AMFB features and the TDNN trained on baseline MFSCs, indicates a 10x faster forward-pass for the TDNN on both regular server CPU (an Intel(R) Xeon(R) CPU E5-2630 v3 @ 2.40GHz)  as well as in GPU (GeForce GTX TITAN X) with about $37$ times fewer parameters than the AMFB-DNN configuration ($37,127,000$ vs. $926,187$). 

Due to the different features, $\sim20\%$ of the DNN architecture parameters vs. $\sim3\%$ for TDNN are accounted for by the input layer alone. Were we to use the same input layer in both architectures, the TDNN will double in size, halving the factor of $37$. However, the more significant reductions in parameters are a consequence of the design decisions of the TDNN, which takes advantage of subsampling and splicing at each layer, allowing a 250-unit output layer to mimic $750$ units while achieving comparable results to the much broader 2048-unit hidden layers in the fully-connected counterpart.

\section{Discussion}
\label{sec:discussion}
The proposed model produces accurate SRT predictions, and --- because the model was trained on speech data without semantic context --- it could potentially generalize to other speech material (as discussed below). 

However, in contrast to existing models, it requires a relatively large amount of training data in the range of $80$ hours, and currently, it is unclear if predictions can be obtained for speech intelligibility across languages. Therefore, it might be challenging to apply this approach to low-resource languages without optimizing the training procedure. In related work targeting listening effort, the method of using phoneme probabilities successfully predicted the listening effort of the German matrix sentence test with fine-tuning using English data \citep{Huber2018}. These findings support the possibility of prediction across languages, at least for phonetically similar languages.

An advantage of the proposed model is that it produces absolute predictions for the SRT, in contrast to some of the baseline models like SII, ESII and STOI, which are normalized using the prediction for a reference condition, in this case, the speech-shaped stationary noise.

Optimally, the corresponding calculations should be carried out on optimized hearing aid hardware. In previous research, we have shown how running at least one feed-forward neural network can be achieved on a hearing aid co-processor \citep{Castro2019}. However, more efficient net topologies such as time-delay neural networks are especially suited for these scenarios as they yield similar or even lower word error rates with much fewer parameters. In this case, just $2.5$\,\% of the AMFB-DNN parameters, allowing for simultaneous measures to compare different processing algorithms.

The fact that both proposed models take advantage of temporal context for predicting speech intelligibility, either via AMFB features and splicing or from the TDNN architecture, provides evidence of such dependencies being essential for predicting SI in the presence of different maskers.

While the model has only been evaluated on matrix sentences with a fixed grammatical structure, these findings may be transferable to more complex speech because the degradation of simple sentences might result in a degradation of conversational speech. Further analysis is needed on more realistic speech data to validate this hypothesis.

For signal changes negatively affecting both the intelligibility of matrix sentences as well as continuous speech (which should be the case for changes due to additive noise and reverberation), we assume consistent model predictions could be obtained (while the absolute SI in terms of recognition score would strongly depend on the speech complexity). 

As pointed out in the previous section, some WER over-/under estimation effects are compared when comparing the model using ASR transcripts and the proposed MTD-based one. However, these effects do not imply the modeling approach with estimated WER is worse as we are interested in predicting human SRTs. 

Furthermore, the MTD measure spreads over enough time to capture subword structures. Even then, we believe, based on our previous observations on large-continuous-read speech recognition frameworks \citep{Castro2019}, that this measure does not exploit the knowledge about the particular speech recognition task, mainly because it is applied to the posteriorgram of context-dependent phonemes rather than the most likely sequence of words. 

Ultimately, the goal is to predict the human error rate to model speech intelligibility, so by focusing on mapping to the WER of an ASR, we might be limiting the predictability of the proposed model.

The model described in this paper has only been investigated for the speech intelligibility of normal-hearing listeners. A model of hearing-impaired listeners could combine techniques such as a degradation of frequency sensitivity on feature level with our model reflecting a specific sensorineural hearing deficit. Both topics, i.e., model predictions for complex speech and modeling speech intelligibility in hearing-impaired listeners, should be addressed in future research.

A simple strategy to consider the hearing loss reflected in a listener's audiogram would be to add frequency-dependent noise to mask the signal properties not accessible to the individual listener. 

The proposed SI model produces accurate predictions for different maskers, listeners, speech enhancement strategies without requiring any reference transcripts. When accounted for other limitations, it could be a crucial component in hearing technology; an important one, needed to overcome before its use outside the lab, is that it has seen the noise during training: In \emph{real-life} applications, the target noise signal is, of course, unavailable. 
However, in related work by \citet{Rossbach2019}, an ASR-based model trained on similar (not identical) noise signals yielded comparable predictions with an RMSE of SRT predictions below 2\,dB. 
Further, a model that combines DNN processing with the M-measure was shown to correlate well with perception when using completely different noise sets for training and testing when an extended training set (8000 hours) was used \citep{Huber2018b}. That study, however, focused on perceived listening effort, which is usually degraded when speech intelligibility is still at 100\,\%, i.e., it remains to be seen if our model generalizes to unknown noise types at low SNRs when using bigger training sets and omitting the decoding step with the HMM. 

The look ahead of the model is two frames, which means the forward pass of a given frame can only be computed $20$\,ms later. This delay is relatively a minor constraint for potential niche applications such as estimating the speech intelligibility resulting from different processing strategies and settings in hearing aids and select the speech enhancement algorithm that maximizes intelligibility in the given acoustic condition. Additionally, the resulting implementation would most likely not be continually computing the speech intelligibility prediction for stability and battery conservation.

While intrusive models are principally not suited for this application scenario, the model introduced in this paper is a step in this direction because the prediction is made without the requirement of clean speech signals or the transcript. In addition to across-language functionality and validating speech intelligibility predictions for hearing-impaired listeners, the next step to be removed towards a reference-free model is to assess the impact of different acoustic environments, with other parameters like reverberation.

A different problem for practical applications is dealing with longer pauses in natural speech. Training material for automatic speech recognition is typically segmented and does not include long silence intervals. An online model would have internal mechanisms for end pointing the utterance according to additional parameters such as SNR and, in more recent developments, an independent voice-activity detection module. A simple peak detector could go a long way to enable or disable the acoustic model to our best knowledge.

Many previous approaches model SI without actually performing the underlying task of speech recognition. Signal characteristics and their degradation can be captured by signal-driven models and do not require recognition or even understanding, shown by the success of established models. However, speech technology recently emerged has enabled human speech perception models, such as the one proposed in this study, albeit only applied to straightforward sentences. Finally, another important future step will be to analyze the underlying deep learning models to determine the cues exploited by the networks to predict speech intelligibility.

\section{Summary}
\label{sec:summary}
This paper explored a modeling approach for speech intelligibility prediction based on ASR without requiring any prior information about the clean speech or the target utterance transcript. 

As a proxy of speech intelligibility, the model predicts the SRT of normal-hearing listeners with similar accuracy to a previous ASR-based model, which requires the transcript of the tested utterance. To overcome this limitation, the proposed model measures the degradation of frame-level phoneme representations obtained from the acoustic model component of the ASR system. Our model outperforms five established baseline models in eight maskers with different types of modulation. 

Two configurations were selected to model the posterior probabilities from the acoustic signal: the first one using amplitude modulation filterbanks with a fully connected feed-forward DNN analogous to the baseline ASR SI model; the second one capturing temporal information using a TDNN on commonly used mel-spectrogram features as an optimized version capable of running on mobile constrained hardware such as hearing aids. 

As the approach was only tested for normal-hearing listeners and with known noise signals, an extension of the model for predicting speech intelligibility of (aided) hearing-impaired listeners in unknown noise should be explored in the future, which would be a significant step towards using this model in-the-loop for real-time optimization in assistive-hearing on mobile hardware.

\section*{Acknowledgments}

This work was funded by the Deutsche Forschungsgemeinschaft (DFG, German Research Foundation) under Germany's Excellence Strategy - Cluster of Excellence Hearing4all EXC 2177/1 Project ID 390895286 and the SFB/TRR 31/3 "The active auditory system," Transfer Project T01.

\bibliographystyle{elsarticle-harv}
\bibliography{refs}

\end{document}